# Collective atom phase controls in photon echoes for quantum memory applications I: Population inversion removal

B. S. Ham

School of Electrical Engineering and Computer Science, Gwangju Institute of Science and Technology
123 Chumdangwagi-ro, Buk-gu, Gwangju 61005, South Korea

bham@gist.ac.kr

(Dated: December 08, 2016)

**Abstract**
Photon echo-based quantum memories demonstrated in rare-earth doped solids over the last decade have solved the major constraint of population inversion in conventional photon echoes by using collective atom phase controls. Both atomic frequency comb and gradient echoes have also made a breakthrough in higher echo efficiency, where conventional photon echo efficiency remains at a few per cent. Here we review, analyze, and discuss the collective atom phase control applied to conventional photon echoes for quantum memory applications to clarify fundamental physics of coherent transients in a three-level system specifically for optical-spin coherence conversion in a controlled double rephasing echoes. Some critical misunderstandings in various protocols are also analyzed, and corrected for near unity echo efficiency under no population inversion.

**Introduction**

Over the last two decades quantum technologies have been employed toward potential applications of quantum cryptography[1-5] and quantum computers[6-10]. However, many technological limitations have been discovered in both areas. Of them are nondeterminacy in single photon- and entangled photon-pair generations, a low detection efficiency in photodetectors, a limited scalability in qubits, and a short (coherence) storage time in quantum memories (qubits). The quantum memory especially has become an essential component for both quantum computers[11,12] and quantum repeaters[13-15] in providing scalability. In that sense both multimode accessibility and ultralong storage time have become the main parameters for determining the functionality of quantum memories.

In light-matter interactions, light absorption strongly depends on the interaction cross section in a matter, where a single atom has extremely small cross section compared with that in an ensemble. To compensate the small cross section in a single atom, the use of an optical cavity is inevitable in sacrificing bandwidth reduction[16-21]. Compared to a single atom-trap storage technique, an ensemble-based one[22-39] gives inherent benefits of single-shot readout, multimode information processing, and frequency division multiplexing. In 2000s Raman scattering and ultraslow light have been applied to quantum memoires in an ensemble medium[33-36], where single-mode information processing and low retrieval efficiency less than 50% have still been the main drawbacks. Although it is a bit impractical, multimode storage in ultraslow light is also possible in an ultradense optical medium[37]. As shown in an electronic chip size gradually shrunken over decades[38], a bulky quantum optical system on an optical table top could also be miniaturized into a quantum chip someday. In that sense, recently demonstrated photon echoes in a nano-cavity may pave the road to future quantum technologies[39].

Photon echoes[40] intensively studied in 1980s and 1990s have given great benefits of all-optical, ultrafast, multimode, and random access characteristics to optical information processing[41-45]. Unlike other ensemble-based quantum memories of Raman scattering and ultraslow light, photon echoes root in coherent transient effects, where inhomogeneous broadening of atoms is an essential requirement. Especially persistent spectral hole-burning rare-earth-doped solids have been intensively investigated due to the benefits of a narrow optical linewidth down to sub-kHz, wide bandwidth greater than GHz, frequency division multiplexing over thousand optical channels, and ultraslow spin decay time as long as a sub-second[46]. Most of all, the time reversible process offers an essential physics for quantum information processing in terms of unitary evolutions. However, ultralow echo efficiency at a few percent has put all potential applications so far on hold.

In spite of the great benefits in all-optical information processing, the photon echo cannot be directly applied for quantum memories due to its inherent population inversion constraint, where the population inversion is an essential step for echo generations in the time-reversed coherence rephasing process. The population inversion, however, induces spontaneous and stimulated emissions resulting in quantum noises. Moreover, the stimulated emission should violate no cloning theorem in quantum information, which state: Any unknown quantum state cannot be duplicated[47]. Thus, there is no way to apply photon echoes directly for quantum information processing except to solve the inherent population inversion constraint. Here we review and analyze collective atom phase controls demonstrated for quantum memories to overcome the population inversion constraint in conventional photon echoes. Then we introduce and discuss the physics of collective atom phase control applied to double rephasing photon echoes, and correct its misuses in recently demonstrated quantum memories in solids. By the way, ultralong quantum memory protocols[48,49] will be discussed elsewhere.

**Review: Solutions to the population inversion constraint**

Figure 1 shows the storage mechanism of the conventional two-pulse photon echoes, where the π optical pulse R rephases all inhomogeneously broadened atoms' phase evolutions triggered by the data pulse D, resulting in a photon echo. For the analysis, the time-dependent density matrix equations, $\dot{\rho}_{ij} (= -\frac{i}{\hbar}[H,\rho] + decay\ terms)$ are numerically solved without any approximations[50]. The following equations are coherence terms of $\dot{\rho}_{ij}$ in a lambda-type, three-level system interacting with two resonant optical fields, obtained by solving the time-dependent Schrodinger equations under rotating wave approximations, $i\hbar|\dot{\Psi}\rangle = H|\Psi\rangle$ ($H$ is interaction Hamiltonian; $\rho = |\Psi\rangle\langle\Psi|$):

$$\frac{d\rho_{12}}{dt} = -\frac{i}{2}\Omega_1(\rho_{11} - \rho_{22}) - \frac{i}{2}\Omega_2\rho_{13} - i\delta_1\rho_{12} - \gamma_{12}\rho_{12}, \quad (1)$$

$$\frac{d\rho_{13}}{dt} = -\frac{i}{2}\Omega_2\rho_{12} + \frac{i}{2}\Omega_1\rho_{23} - i(\delta_1 - \delta_2)\rho_{13} - \gamma_{13}\rho_{13}, \quad (2)$$

$$\frac{d\rho_{23}}{dt} = -\frac{i}{2}\Omega_2(\rho_{22} - \rho_{33}) + \frac{i}{2}\Omega_1\rho_{21} + i\delta_2\rho_{23} - \gamma_{23}\rho_{23}. \quad (3)$$

Here $\Omega_1$ is the Rabi frequency of the resonant optical field (related with photon echoes) between the ground state $|1\rangle$ and the excited state $|2\rangle$, $\Omega_2$ is the Rabi frequency of the resonant control field between the auxiliary ground state $|3\rangle$ and the excited state $|2\rangle$, and $\delta_1$ ($\delta_2$) is the atom detuning from the resonance field $\Omega_1$ ($\Omega_2$). Instead of using Maxwell-Bloch equations as done in many photon echo studies with appropriate approximations[51-54], we focus on the coherence evolution of individual atom phases in time domain without approximations. Although complete light-matter interactions can be solved by combining both density matrix equations and Maxwell-Bloch equations, a complete solution has been limited to the short time scale of picoseconds, due to the limitations of computing resources[55]. For all numerical simulations, nine total time-dependent density matrix equations including equations (1)~(3) are calculated without any approximations.

For photon echoes, the medium must be inhomogeneously broadened, and the data pulse spectrum must be within this broadening. The optical pulse $\Omega$ is assumed to be monochromatic and a square pulse. For a two-level system in Fig. 1a, we set $\Omega_2$=0 and $\gamma_{3j}$= 0 (j=1,2). Initial conditions are $\rho_{ij}$=0, except for $\rho_{11}$=1. Figure 1b shows a normal photon echo simulation. As shown in Fig. 1c,e,g, inhomogeneously broadened atoms excited by D (π/2 pulse area) induces a time-dependent phase grating in a spectral domain of the ensemble medium. The modulation frequency of the phase grating at a given time t is determined by $1/(t - t_D)$. By the π pulse R, this phase grating gains a π phase shift (see Fig. 1d): $\rho(t) \rightarrow \rho(t)^*$. Thus, this phase grating becomes the storage mechanism in two-pulse photon echoes.

When the rephasing pulse R is divided into two time-delayed pulses, conventional three-pulse photon echo scheme is satisfied[56]. In Fig. 1e,f, the phase grating just before R at t=10.0 μs is now converted into a population grating by the first half R at t=10.05 μs satisfying the pulse area of π/2. In Fig. 1g,h (blue curves), the modulation frequency is 200kHz (=1/5 μs) at $t = 10.0\ \mu s$: $t - t_D = 5\ \mu s$. Then the population grating turns out to be its original phase grating by the second half R via rephasing process with a π phase shift (the color swapping across R). Here, the important physics is that the coherence conversion between phase and population gratings is reversible. Thus, the population grating becomes the storage mechanism of not only conventional three-pulse photon echoes[41-45,56], but also quantum memories as shown in atomic frequency comb (AFC) echoes[23,24,28,52]. The extremely weak retrieval efficiency in photon echoes roots in echo reabsorption governed by Beer's law, which is another constraint to be solved for quantum memory applications (will be discussed elsewhere).

The first trial to fix the population inversion constraint in photon echoes was done by Swedish and Russian groups in 2001 (ref. 30) followed by Korean and Russian groups in 2003 (ref. 31), in the name of controlled reversible inhomogeneous broadening (CRIB) in a three-level system. In the CRIB echoes in Fig. 2, the counter-propagating control pulses (C1 & C2), whose pulse area is $\pi$ each, causes a coherence inversion *with a $\pi$ phase shift* $[\rho(t) \rightarrow -\rho(t)^*]$ between symmetrically detuned atom pairs at $\pm\delta_j$ via **k**-dependent opposite Doppler shifts in an atomic vapor, resulting in a photon echo without population inversion (see Fig. 2c). Figure 2d~f shows swapping in the coherence evolution direction by C pulses. In Fig. 2f, the opposite Doppler shifts are visualized with symmetrically detuned atoms ($\pm\delta$=50 kHz) under non-zero phase decay rate, where a $+\delta$ detuned atom plays as a $-\delta$ detuned atom in phase evolution by the control pulses. This coherence inversion in CRIB echoes is definitely different from the rephasing mechanism in photon echoes (Fig. 1d), where the rephasing never changes the coherence evolution direction of each atom in a Bloch vector plane (Supplementary Fig 1). The CRIB mechanism in Fig. 2, however, needs modification to apply for a solid medium due to no Doppler effects, resulting in $\rho(t) \rightarrow -\rho(t)$ (will be discussed in Fig. 4) (ref. 31). Here the population transfer to the spin state $|3\rangle$ affects storage-time extension up to spin $T_2$, which is much longer than the optical one[24,30,31,34-36].

Few years later in 2006, the idea of CRIB had developed into gradient echoes in a two-level solid medium by an Australian group[57]. In the gradient echoes, a reverse dc electrode pair plays the role of the control pulses in the CRIB technique in Fig. 2, resulting in an echo without population inversion. For this the medium must be persistent spectral hole burning and optically dense for spectral expansion by the gradient electric field. Unlike conventional photon echoes governed by Beer's law resulting in severe echo reabsorption, CRIB-based modified photon echoes offer near unity echo efficiency[30,58]. The dc electrode may limit potential applications of gradient echoes in a bulk medium due to the electrode size necessary for linear gradient field generation.

The second trial to fix the population inversion constraint in photon echoes was done by a Geneva group in 2008 using AFC echoes[23]. The AFC technique is based on atom population grating in a persistent spectral hole-burning medium composed of at least three energy levels. Unlike the three-pulse photon echoes whose grating is formed by consecutive two pulses, a repeated weak two-pulse train renders the population grating sharper, while lowering the absorption efficiency[52]. Such a sharper population grating however, induces a lower retrieval efficiency due to lesser absorption. The weaker AFC echoes is of course compensated by the coherence accumulation (phase grating induced AFC) obtained by the repeated weak two-pulse train (Supplementary Fig. 2). Thus, efficient AFC echoes can be obtained via maximizing rephasing efficiency, while minimizing echo resorption[59]. Moreover, the accumulated coherence of the population grating in AFC allows multiple storages (actually readouts) as shown in the Inset of Supplementary Fig. 2c (ref. 60). Unlike three-pulse echoes[41-45,56], the storage time of AFC echoes is predetermined by the weak two-pulse delay time used for AFC, which roots in the phase grating in two-pulse photon echoes (Fig. 1g). Retrieval efficiency in bare AFC echoes without an optical cavity, however, is still far less than 50%. The physics of AFC echoes has been intensively studied recently[61].

The third technique for solving the population inversion problem in photon echoes was presented by a Korean group in the name of controlled double rephasing (CDR) echo in 2011 (ref. 62). Because each $\pi$ pulse in a two-level system induces a population inversion, double $\pi$ pulses should remove the population inversion constraint. To work with this protocol, there are two major requirements to be satisfied. *Firstly, the first echo generated by the first $\pi$ optical pulse must be killed (erased or silent) not to affect the second echo, and secondly, the second echo must be emissive in collective atom coherence.* For the first requirement a French group introduced a silent echo concept with on-demand phase mismatching[63]. Because a photon echo is a direct result of a nonlinear macroscopic (or collective) coherent transient effect, a simple way to destroy echo formation is to add controlled phase turbulence or to violate the phase matching condition. Such silent echoes have also been investigated by using Stark[64,65] and magnetic[66,67] effect-based phase turbulence in a single rephasing scheme. According to the phase matching condition, two-pulse photon-echo wave vector $\mathbf{k}_{e1}$ is given by $\mathbf{k}_{e1} = 2\mathbf{k}_R - \mathbf{k}_D$, where $\mathbf{k}_i$ stands for the wave vector of a pulse $i$. If a backward rephasing pulse ($\mathbf{k}_R = -\mathbf{k}_D$) is applied, the echo wave vector becomes $\mathbf{k}_{e1} = -3\mathbf{k}_D$. Then, the phase mismatch between $\mathbf{k}_D$ and $\mathbf{k}_{e1}$ occurs due to different magnitude of the wave vector: The wavelength-dependent refractive index should not satisfy the phase matching condition in a bulk medium: $|k_{e1} - k_D|L = \frac{2\pi}{\lambda}|n(3\omega) - n(\omega)|L \gg \pi$. Thus, the echo e1 becomes silent. The resulting macroscopic phase mismatch, however, does not affect the individual phase evolutions. Therefore, the second echo e2 by the second π optical pulse RR is generated satisfying the phase matching condition under no population inversion: $\mathbf{k}_{e2} = 2\mathbf{k}_{RR} - \mathbf{k}_{e1} = \mathbf{k}_D$, if $\mathbf{k}_{RR} = \mathbf{k}_R = -\mathbf{k}_D$. However, this echo e2 in a simple double rephasing scheme like in ref. 63 shows an absorptive characteristic as a direct result of the double rephasing (will be discussed in Fig. 5). This is why the CDR scheme is needed (will be discussed in Fig. 6).

**Analysis and Discussions**
   **A. Atom phase control in two-pulse photon echoes: Controlled photon echoes**
   To discuss CDR echoes, we start with optical Rabi flopping in a resonant Raman system. Here, our interest is how the Rabi flopping affects coherence. Coherence conversion between optical and spin states has already been analyzed in ultraslow light-based quantum memories as a key mechanism[34-37]. In Fig. 3, control Rabi flopping-affected optical coherence is discussed for a resonant Raman system. For this, a two-level system is taken as a reference: $\Omega_C = 0$; $\gamma_{23} = \gamma_{13} = 0$. For optical excitations with Rabi frequency $\Omega_D$ in a two-level system, the state vector $|\Psi(t)\rangle_D$ is described by:

$$|\Psi(t)\rangle_D = \cos\left(\frac{\Omega_D t}{2}\right)|1\rangle + i\sin\left(\frac{\Omega_D t}{2}\right)|2\rangle, \tag{4}$$

where its coherence term $(c_1 c_2^*)$ is denoted by $-i\cos\frac{\Omega_D t}{2}\sin\frac{\Omega_D t}{2}\left(= -\frac{i}{2}\sin\Omega_D t\right)$, resulting in a $\Omega_D$ Rabi oscillation (first parts of Fig. 3c-f in time domain). In a three-level Raman system of Fig. 3a, the state vector $|\Psi(t)\rangle_R$ with optical Rabi frequencies $\Omega_D$ and $\Omega_C$ is described by:

$$|\Psi(t)\rangle_R = \left[\frac{\left(\Omega_C^2 + \Omega_D^2 \cos\left(\frac{\Omega t}{2}\right)\right)}{\Omega^2}\right]|1\rangle + i\frac{\Omega_D}{\Omega}\sin\left(\frac{\Omega t}{2}\right)|2\rangle + \frac{\Omega_D \Omega_C}{\Omega^2}\left[\cos\left(\frac{\Omega t}{2}\right) - 1\right]|3\rangle, \tag{5}$$

where $\Omega_D$, $\Omega_C$, and $\Omega$ are the Rabi frequencies of D, C, and Raman pulses, respectively, and $\Omega = \sqrt{\Omega_D^2 + \Omega_C^2}$. The Raman state vector in equation (5) oscillates twice slower than equation (4) at $4\pi$ of $\Omega$ (second parts of Fig. 3b-f). Such a $4\pi$ Raman coherence oscillation has already been experimentally demonstrated in resonant Raman echoes[68]. If $\Omega_C \gg \Omega_D$, the control pulse $\Omega_C$ (~$\Omega$) in equation (5) becomes a dominant factor:

$$|\Psi(t)\rangle_R \cong |1\rangle + \frac{\Omega_D}{\Omega}\left[\cos\left(\frac{\Omega t}{2}\right)|3\rangle + i\sin\left(\frac{\Omega t}{2}\right)|2\rangle\right]. \tag{6}$$

In other words the C-induced (Raman) Rabi flopping inverts the system coherence at every $2\pi$ of $\Omega T$: $|\Psi(t+T)\rangle_R = -|\Psi(t)\rangle_R$. This control Rabi flopping resembles the CRIB case of Fig. 2e without Doppler effects (will be discussed in Fig. 6). Here the two cases in Fig. 3 are completely independent as shown $Re\rho_{13}=0$ in Fig. 3c.
   With the D pulse only for a direct transition, the period of optical coherence $\rho_{12}$ matches the population oscillation period (e.g. for $\rho_{22}$: $2\pi$ oscillation period in Fig. 3c,d). The optical coherence $\rho_{12}$ in the Raman system, however, oscillates at $4\pi$ basis, resulting in two oscillations total. In the Raman system, the excited state population $\rho_{22}$ keeps the same period as in the direct excitation (dotted curve in Fig. 3d), while others ($\rho_{11}$ and $\rho_{33}$) coincide with the Raman coherence $\rho_{13}$ (red curve in Fig. 3c). The population difference ($\rho_{11} - \rho_{22}$) in Fig. 3f also coincides with $\rho_{13}$: $\rho_{11} + \rho_{22} + \rho_{33} = 1$. As denoted in equation (1), the optical coherence $\rho_{12}$ in a three-level system is affected by both Raman coherence ρ₁₃ and population difference $\rho_{11} - \rho_{22}$. As briefly mentioned in equation (6), Fig. 3e,f prove this statement. The analytic solution of optical coherence $\rho_{12}(t)$ $(= c_1 c_2^*)$ from equation (5) is

$$\rho_{12}(t) = -i \frac{\Omega_D \left[\Omega_C^2 + \Omega_D^2 \cos\left(\frac{\Omega t}{2}\right)\right] \sin\left(\frac{\Omega t}{2}\right)}{\Omega^3}. \quad (7)$$

As discussed in Fig. 3, equation (7) also confirms the 4π oscillation period of the control pulse C in the optical coherence ρ₁₂.

What happens if the pulses D and C are temporally separated in the resonant Raman scheme of Fig. 3b? Figure 4 is for an extension of Fig. 3 for the delayed Raman scheme, which is prerequisite for CDR echoes. In the light-matter interactions, the time delay between the resonant Raman pulses must be shorter than the inverse of the optical inhomogeneous width[69]. This is the direct result of macroscopic coherence as appeared in free induction decay (FID). However, in the coherent transients such as photon (spin, or Raman) echoes, this rule does not apply anymore, because rephasing is based on individual atom coherences, governed by optical homogeneous decay time.

The collective (overall) Raman coherence excitation ρ₁₃ in a delayed scheme of Fig. 4a is zero all the time. The collective (overall) optical coherence ρ₁₂ decays as a function of the inverse of the optical inhomogeneous broadening [1/(510,000π)=0.6 μs]: optical FID. However, individual atoms are affected by the optical homogeneous decay time, which is set ∞ for an ideal system here (Fig. 4b). We also set zero spin dephasing for simplicity. As mentioned in Fig. 3, the optical coherence ρ₁₂ of individual atoms oscillates twice slower in the Raman system (Fig. 4b,c). Thus, a 2π control pulse C induces a coherence inversion (ρ₁₂ → −ρ₁₂) as discussed in equations (5) and (6) (Fig. 4d). This means that a 2π (4π) control pulse applied to photon echoes results in an absorptive (emissive) echo as shown in Fig. 4e (refs. 48,62,68). Figure 4f represents that the single 4π pulse area of C in Fig. 4e can be divided into two control parts, C1 (π) and C2 (3π) (refs. 70,71). In this case only matter is spin dephasing[70]. Here the controlled photon echoes in Fig. 4f, however, still keep the population inversion constraint. The solution will be discussed in Fig. 6 in the name of CDR echo.

Although 2π (or π−π) control C in Fig. 4g offers the same coherence inversion as in CRIB of Fig. 2e, the rephased coherence evolution direction is opposite each other, resulting in an absorptive echo in Fig. 4e: Non-Doppler vs. Doppler. Thus, unlike the π−π control pulse sequence in a Doppler medium[30], the same control pulse sequence applied to the controlled AFC echo in ref. 24 is a mistake. The reason for experimental observations of the controlled AFC echoes, however, may be due to the Gaussian spatial distribution in a transverse mode of the light pulses as well as Beer's law-dependent absorption strength in an axial mode, resulting in all possible pulse areas. Moreover, rare-earth doped solids governed by crystal fields allow mixed atomic transitions[72], resulting in an imperfect population transfer between the excited and auxiliary states[73]. This partially violates the atom phase recovery condition of 4nπ of the control Rabi flopping discussed in Fig. 4h (ref. 71). Thus, any pulse area should contribute to photon echo generations. Only matter is echo efficiency: see Fig. 3(a) of ref. 74. To solve the absorptive echo problem in Fig. 4g as well as in ref. 24, another 2π control pulse is needed as shown in Fig. 4h: A full analytical expressions have been discussed elsewhere[75]. With counter-propagating C1 (π) and C2 (3π), the echo direction $\mathbf{k}_E$ can be controlled to be opposite with respect to $\mathbf{k}_D$ ($\mathbf{k}_E = -\mathbf{k}_D + \mathbf{k}_{C1} + \mathbf{k}_{C2}$), showing ideal, near perfect echo efficiency[30,31,70].

### B. Doubly rephased photon echoes

Now we analyze a doubly rephased two-pulse photon echo without a control pulse in Fig. 5: $\Omega_2=0$ and $\Gamma_{23}=\gamma_{23}=\Gamma_{31}=\gamma_{31}=0$. Initially all atoms are in state $|1\rangle$: $\rho_{11}=1$. Figure 5b shows overall coherence $Im\rho_{12}$, where the first echo e1 and the second echo e2 are denoted. A part of atoms on the ground state $|1\rangle$ are excited by a weak (or quantum) data pulse D and start to freely evolve as a function of time (unitary evolution): $\Psi(r,t) = \Psi(r)e^{\pm i\delta_j t}$ in the rotating wave approximation. For simplicity let $t_D = 0$ and $t_R - t_D = T$, where $t_z$ represents the arrival time of pulse z. The detuning $\pm \delta_j$ is for a symmetrically detuned $\pm j^{th}$ atom pair across the line center of N contributed atoms in an optical inhomogeneous broadening $\Delta$: $\Delta = \frac{1}{N}\sum_{j=1}^{2N} \delta_j \rightarrow Gaussian\ shape$. As shown in Fig. 5c,d, the first $\pi$ pulse R (c-d) rephases the system coherence with a $\pi$ phase shift at $t = t_R = T$, and individual atom phase evolutions continue in $t'$: $e^{\pm i\delta_j t} \rightarrow e^{\mp i\delta_j T} e^{\pm i\delta_j t'} = e^{\pm i\delta_j(t'-T)}$; $t' = t - t_R$. At $t = t_{e1} = t_R + T = 2T$, the first echo e1 can be generated in silence under on-demand phase mismatch[63], or by adding phase turbulence[64-67,74].

The second $\pi$ pulse RR (e-f) in Fig. 5c,d arrives after e1 at $t = t_{RR} = t_{e1} + T' = 2T + T'$ (or $t' = T + T'$), rephases the system again based on the first echo e1, and the atom phase evolutions continue in $t''$: $e^{\pm i\delta_j(t'-T)} \rightarrow e^{\mp i\delta_j(T+T'-T)} e^{\pm i\delta_j t''} = e^{\pm i\delta_j(t''-T')}$; $t'' = t - (2T + T')$. Thus, at $t = t_{e2} = 2(T + T')$, the rephased atoms are coherently lined up for echo e2 under no population inversion. However, the echo e2 cannot be radiated from the medium because its macroscopic coherence is absorptive like the data D as shown in Fig. 5b,d (mark 'x'). This absorptive echo e2 is obvious, where the double rephasing results in a $2\pi$ phase shift (i.e., no phase shift): $\rho(t) \xrightarrow{R1} \rho(t')^* \xrightarrow{R2} \rho(t'')$. As discussed in Fig. 4g for the controlled AFC (ref. 24), the doubly rephased photon echo in ref. 63 is also absorptive. The echo observation in ref. 63 is also due to nonuniform pulse area applied to each atom resulting from a Gaussian pulse shape in a transverse spatial mode perpendicular (x and y axis) to the beam propagation direction (z axis) as well as Beer's law-dependent absorption along the longitudinal axis. To fix the absorptive echo in 5b as well as in ref. 63, the CDR echo is introduced in Fig. 6.

### C. dc Stark echoes in the double rephasing scheme

The dc Stark echo is another quantum memory protocol applied to double rephasing photon echoes with dc Stark-induced phase turbulence for the silent echo formation[74,76]. The dc Stark control in double rephasing scheme has an advantage in removing drawbacks in gradient echoes[26,57,58], where a persistent spectral holeburning-, ultradense-, and bandwidth-limited optical medium has been the prerequisite conditions. The dc Stark effect was first observed in two-pulse spin echoes half a century ago[64], where spin echoes is the magnetic version of photon echoes[77]. Unlike the gradient echoes, the atoms do not need to be initially prepared for a narrow spectral antihole, but the length limitation of the electrodes should still be remained as a constraint. Supplementary Fig. 3 shows the dc Stark echo schematics, where two unbalanced dc Stark pulses (DC1 and DC2) are inserted across the first rephasing pulse R1, and followed by the second rephasing pulse R2. Here the "unbalanced" stands for silencing the first echo, where DC2 cannot come before the first echo E1. Due to the $\pm\Delta\omega$ Stark splitting symmetry (Supplementary Fig. 3), the excited atoms (spins) by D are divided into two groups, resulting in fast and slow phase evolutions. Thus, the phase accumulations between two groups are interfered and cancelled out only at a specific condition for no echo generation[64,74,76]: $\rho_{12} = \sum \rho_{12}^{(j)}(e^{+i\Delta\omega\tau} + e^{-i\Delta\omega\tau}) \propto \cos(\Delta\omega\tau)$. Thus, the silent echo condition for the dc Stark shift by DC1 is $\Phi_{DC1} = \Delta\omega\tau = (2n-1)\pi/2$.

Because the role of the second dc Stark pulse DC2 is to compensate the DC1-induced phase shift $\Phi_{DC1}$, two conditions must be satisfied for DC2. *First, each Stark induced phase shift must be equal*: $\Delta\omega_1\tau_1 = \Delta\omega_2\tau_2$. The subscript 1 (2) indicates for DC1 (DC2). *Second, the second echo must be emissive to be radiated as discussed in Section B*. If exactly the same Stark fields are applied as in ref. 76, however, the D-excited atom coherence evolution turns out to be absorptive echo generation:

$e^{\pm i\delta_j t} \xrightarrow{DC1} e^{\pm i\delta_j t}(e^{-i\Delta\omega_1\tau_1} + e^{+i\Delta\omega_1\tau_1}) \xrightarrow{R1} e^{\pm i\delta_j(t'-T)}(e^{+i\Delta\omega_1\tau_1} + e^{-i\Delta\omega_1\tau_1})$
$\xrightarrow{DC2} (e^{+i\Delta\omega_1\tau_1} + e^{-i\Delta\omega_1\tau_1} + e^{+i\Delta\omega_2\tau_2} + e^{-i\Delta\omega_2\tau_2}) \cdot e^{\pm i\delta_j(t'-T)} \xrightarrow{R2} e^{\pm i\delta_j(t''-T')}$; $t' = t - T$;

$t'' = t - t_{R2} = t - (2T + T')$. This is same as the doubly rephased echo e2 in Fig. 5d, which is absorptive. Here, the detuning signs applied to the atoms by the dc Stark splitting are predetermined to each atom in a solid, i.e. DC1 and DC2 split the same atom groups without intermixing[78]. For Supplementary Fig. 3b, $\Delta\omega_1 \neq \Delta\omega_2$ must be satisfied. Thus, the second requirement of dc Stark echoes is not satisfied in ref. 76.

On the other hand in ref. 74, the gradient echo concept is used with reversed electric field polarization between DC1 and DC2. Thus, the interaction with two dc Stark fields results in not only phase turbulence compensation, but also a π phase shift as discussed for CRIB in Fig. 2: $[\rho(t) \rightarrow -\rho(t)^*]$. Thus, echo E2 becomes emissive in ref. 74, and the dc Stark echo protocol works.

### D. Controlled Double Rephasing (CDR) echoes

To remove the population inversion constraint, the CDR echo has been proposed[62]. CDR echo also solves the absorptive echo problems in a double rephasing scheme of Fig. 5. Figure 6 shows the CDR echo calculations. The main purpose of the CDR scheme is to remove the population inversion constraint in conventional photon echoes, and now it has become a powerful tool to solve the absorptive photon echo dilemma in various modified photon echo schemes for quantum memory applications[24,63,76]. Unlike the π–3π control pulse sequence in a single rephasing scheme of Fig. 4f (ref. 70), we need a π–π control pulse sequence to make an emissive echo in a double rephasing scheme (e2 in Fig. 6b). As shown in Fig. 6c,d, the individual atom coherence of the doubly rephased echo e2 at point 'f' experiences a complete coherence inversion by the C1(π)–C2(π) (or 2π C), and reaches at point 'h' through the point 'g,' which is zero optical coherence: $e^{\mp i\delta T'} \rightarrow -e^{\mp i\delta T'}$: $T' = t_{e1} - t_R$. Thus, coherence evolution after C2 is denoted by $-e^{\pm i\delta(t''-T')}$, where $t'' = t - (2T + T' + T_C)$, $T = t_R - t_D$, and $T_C = t_{C2} - t_{C1}$. This means that the coherence $\rho_{12}(t)$ becomes an emissive echo e2 at $t = 2(T + T') + T_C$ (mark 'x' in Fig. 6d). It should also be noted that the zero optical coherence at point 'g' offers a storage extension benefit by $T_C$, which is limited by spin phase decay time[24,30,31,61,62,70,71]. The storage time extension up to spin population decay time is beyond the scope of this article, and will be discussed elsewhere[48,49]. The control pulse set of C1 and C2 can be positioned after R but before e1[75].

For the silent echo e1 in the CDR echoes, the on-demand phase-shift control analyzed in refs. 63,76 can be applied. The silent echo e1 and emissive echo e2 can also be obtained by using unbalanced ac Stark shifts[79]. As presented in the CDR scheme[62], counter-propagating C1 (π) and C2 (π) can control the echo direction $\boldsymbol{k}_{e2}$ to be opposite $\boldsymbol{k}_D$ ($\boldsymbol{k}_{e2} = -\boldsymbol{k}_D + \boldsymbol{k}_{C1} + \boldsymbol{k}_{C2}$), resulting in an ideal near perfect echo efficiency[30,31,70]. Here rephasing pulse has nothing to do with the four-wave mixing processes as experimentally demonstrated[70].

**Conclusion**

Various modified photon echo protocols demonstrated for quantum memory applications were reviewed, analyzed, and discussed to give a better understanding of the collective atom phase control for inversion-free photon echoes. Controlled coherence conversion by a control Rabi pulse pair resonant between an excited state and an auxiliary spin state was discussed for atom phase control to solve the absorptive echoes analyzed in several modified photon echoes such as controlled AFC in a single rephasing scheme and dc Stark echoes in a double rephasing scheme. The experimental echo observations in these schemes might be due to the Gaussian pulse shape in a transverse spatial mode as well as Beer's law-dependent absorption strength in the axial mode, resulting in all different kinds of pulse areas to each atom. To solve the absorptive echo problem in the double rephasing schemes, a CDR echo protocol was introduced, analyzed and discussed for inversion-free emissive photon echoes. The CDR echo protocol also offers benefits of near perfect echo efficiency via nondegenerate four-wave mixing processes. For real quantum memory applications in the modified echoes, firstly, the spatial transverse mode of the optical pulses should be made uniform, and secondly, the Beer's law-dependent absorption strength should be flattened (will be discussed elsewhere).

**Acknowledgment**
This work was supported by ICT R&D program of MSIP/IITP (1711028311: Reliable crypto-system standards and core technology development for secure quantum key distribution network) and GIST-Caltech Program in 2016. BSH thanks for discussions with T. Zhong, J. M. Kindem, J. Bartholomew and A. Faraon at Caltech.

62. Ham, B. S. Atom phase controlled noise-free photon echoes. arXiv:1101.5480 (2011); ibid, arXiv:1109.5739 (2011).
63. Damon, V., Bonarota, M., Louchet-Chauvet, A., Chanelière, T. & Le Gouët, J-L. Revival of silenced echo and quantum memory for light. *New. J. Phys.* **13**, 093031 (2011); ibid, arXiv:1104.4875 (2011).
64. Mims, W. B. Electric field effects in spin echoes. *Phys. Rev.* **133**, A835-A840 (1964).
65. Meixner, A. J., Jefferson, C. M. & Macfarlane, R. M. Measurement of the Stark effect with subhomogeneous linewidth resolution in $Eu^{3+}$:$YAlO_3$. *Phys. Rev. B* **46**, 5912-5916 (1992).
66. Wang, Y. P., Boye, D. M., Rives, J. E. & Meltzer, R. S. Modulation of photon echo intensity by pulsed non-uniform magnetic fields. *J. Lumin.* **45**, 437-439 (1990).
67. Hetet, G. et al. Photon echoes generated by reversing magnetic field gradients in a rubidium vapor. *Opt. Lett.* **33**, 2323-2325 (2008).
68. Ham, B. S., Shahriar, M. S., Kim, M. K. & Hemmer, P. R. Spin coherence excitation and rephasing with optically shelved atoms. *Phys. Rev. B* **58**, R11825-R11828 (1998).
69. Ham, B. S., Hemmer, P. R. & Shahriar, M. S. Efficient phase conjugation iva two-photon coherence in an optically dense crystal. Phys. Rev. A **59**, R2583-R2586 (1999).
70. Hahn J. & Ham, B. S. Rephasing halted photon echoes using controlled optical deshelving. *New J. Phys.* **13**, 093011 (2011).
71. Ham, B. S. Control of photon storage time using phase locking. *Opt. Exp.* **18**, 1704-1713 (2010).
72. Ellitt, R. J. Crystal field theory in the rare earths. *Rev. Mod. Phys.* **25**, 167-169 (1953).
73. Ham, B. S. A contradictory phenomenon of deshelving pulses in a dilute medium used for lengthened photon storage time. *Opt. Exp.* **18**, 17749-17755 (2010).
74. McAuslan, D. L. et al. Photon-echo quantum memories in inhomogeneously broadened two-level atoms. *Phys. Rev. A* **84**, 022309 (2011).
75. Rhamatt and Ham, B. S. Analysis of controlled coherence conversion in a double rephasing scheme of photon echoes for quantum memories. arXiv:2016.02167 (2016)
76. Arcangeli, A., Ferrier, A. & Goldner, Ph. Stark echo modulation for quantum memories. Phys. Rev. A**93**, 062303 (2016).
77. Hahn, E. L. Spin echoes. *Phys. Rev.* **80**, 580-594 (1950).
78. Ludwig, G. W. & Woodbury, H. H. Splitting of electron spin resonance lines by an applied electric field. *Phys. Rev. Lett.* **7**, 240-241 (1961).
79. Ham, B. S. A controlled ac Stark echoes for quantum memories. arXiv:2016.02193 (2016).

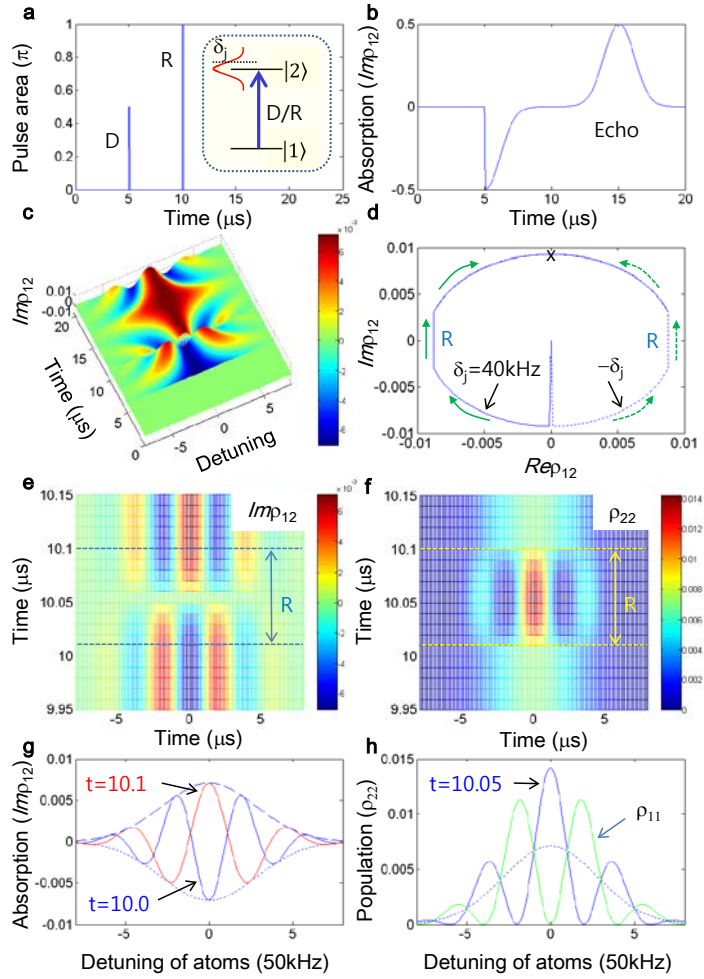

**Figure 1| Storage mechanism in photon echoes. a**, A schematic of two-pulse photon-echo. The pulse arrival time for D (R) is t=5 (10) with 0.1 μs pulse duration. All decay rates are zero. Inset: Energy levels of inhomogeneously broadened atoms (FWHM: 340 kHz). **b**, Numerical simulations for **a**. Photon echo appears at t=15. A 2D color map of **c**, coherence $Im(\rho_{12})$ and **d**, Bloch vector diagram for a detuned atom pair at $\delta_j = \pm 40\ kHz$. The mark ′x′ is for echo timing. **e, f**, Details of **c** for 10.00<t≤10.10. **g**, Details of **c** for phase grating at t=5.1 (dotted, D), t=10.0 (blue, before R), t=10.1 (red, after R), and t=15.0 (dashed, echo). **h**, Details of **c** for population grating ($\rho_{22}$) at t=5.1 (dotted, D), t=10.05 (blue, middle of R). The green curve is for $\rho_{11}$ at t=10.05 (middle of R). In the programmig the step of time increment is 0.01μs. Thus R pulse actually turns on at t=10.01 μs for 10.00<t≤10.10. The time unit is μs.

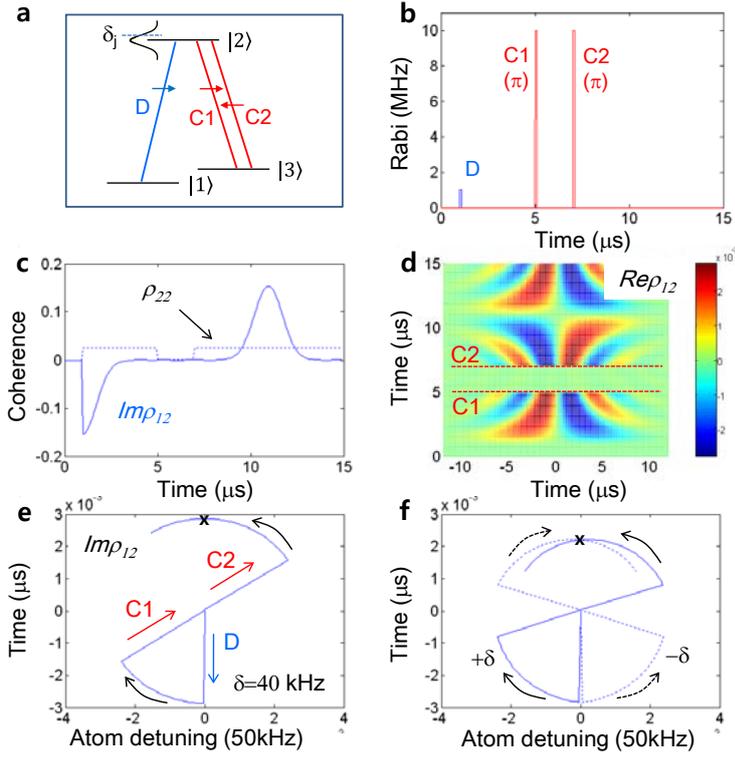

**Figure 2| Controlled reversible inhomogeneous broadening (CRIB) echoes. a**, **b**, Schematics of CRIB echoes. **c**, Numerical simulations of CRIB echo. **d**, 2D picture of **c** for $Re\rho_{12}$. **e**, Bloch vector of a detuned atom coherence evolution: $\delta=40$ kHz. **f**, Bloch vector model for a symmetrically detuned atom pair: $\delta=\pm 50$ kHz for $\Gamma_{21}=\Gamma_{23}=5$ kHz. The optical inhomogeneous width is 510 kHz (FWHM), where 10 kHz-121 groups are used for the calculations. The mark 'x' in **e** and **f** denotes the echo timing. All decay rates are zero, otherwise specified.

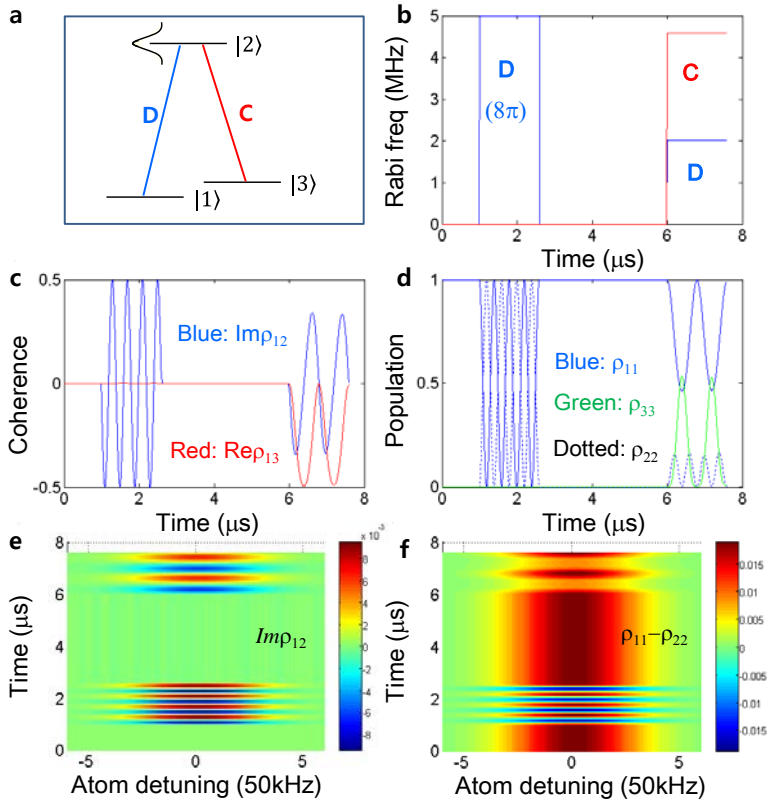

**Figure 3| Direct and indirect coherence excitations. a**, **b**, Schematics of atom-light interactions for direct (D only) and indirect (D and C; resonant Raman) coherence excitations. The pulse area of D and Raman is $8\pi$ each. **c**, **d**, Overall coherence and population oscillations for **b**. A 2D color map of **e**, $Im\rho_{12}$ in **c** and **f**, $\rho_{11}-\rho_{22}$ in **d** for all atoms. The optical inhomogeneous broadening is 300 kHz (FWHM). Initially all $\rho_{ij}=0$, except for $\rho_{11}=1$.

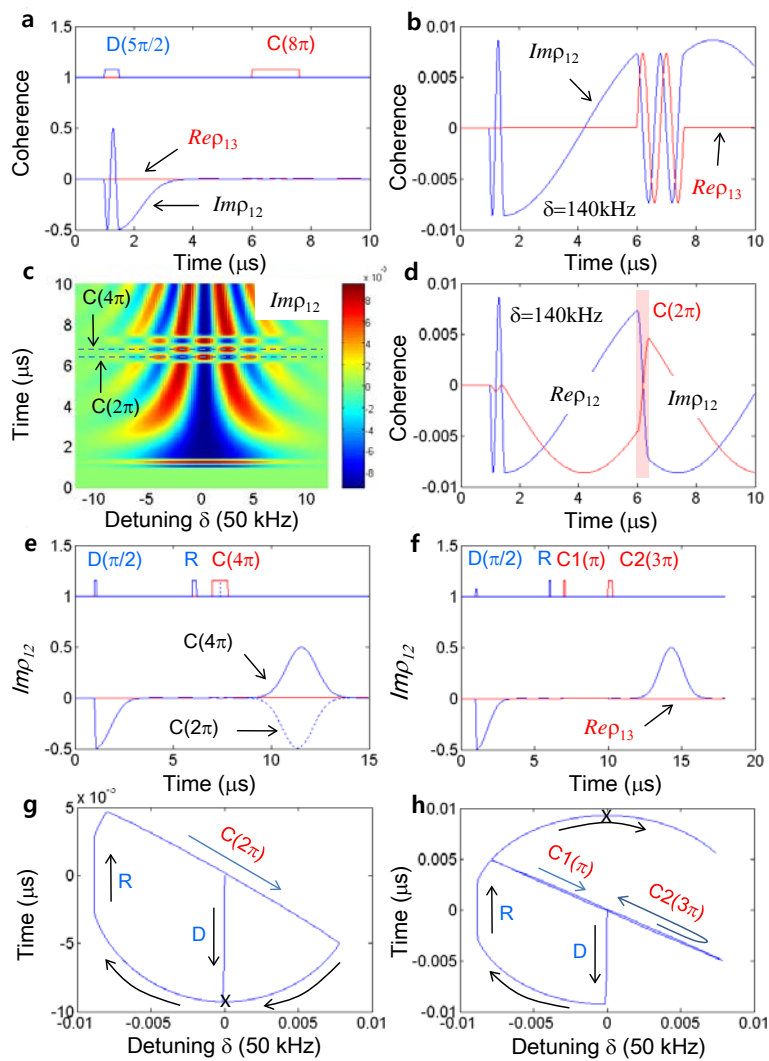

**Figure 4| Atom phase control in delayed resonant Raman system. a**, Time-delayed light-matter interactions for Raman in Fig. **3a**. The pulse area of D and C is $5\pi/2$ and $8\pi$, respectively. **b**, Coherence evolution of a detuned atom in **a**. **c**, A 2D color map of **a**. **d**, Coherence inversion by a $2\pi$ control pulse C in **c**. **e**, Atom phase control by C. **f.** Controlled photon echo with C1($\pi$) and C2($3\pi$). **g, h**, Bloch vector models for a detuned atom with coherent control C. The mark 'x' is for the echo timing. Optical inhomogeneous broadening is 300 kHz (FWHM). All $\rho_{ij}(t=0)=0$, except for $\rho_{11}(t=0)=1$. All decay rates are zero.

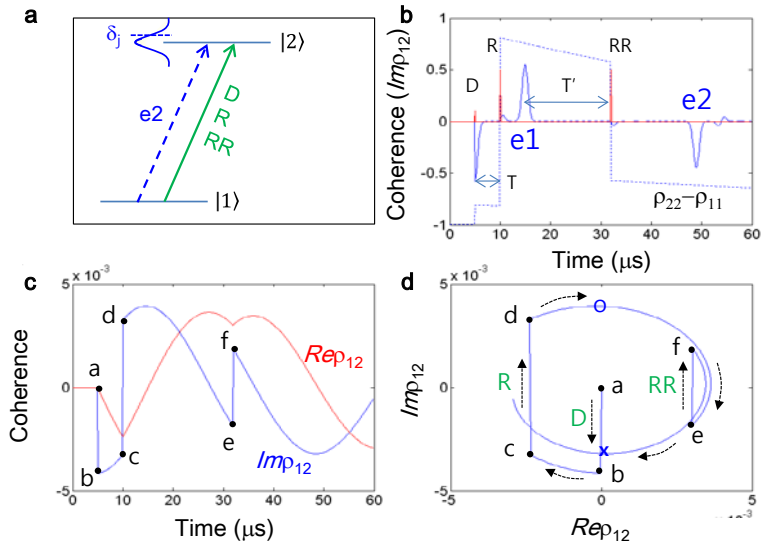

**Figure 5| Double rephasing (DR) in photon echoes. a**, A schematic of DR. **b**, Numerical simulations of DR echo: $t_D=5$; $t_R=10$; $t_{RR}=32$; $t_{e1}=2t_R-t_D=15$; $t_{e2}=2t_{RR}-t_{e1}=2t_{RR}-2t_R+t_D=49$ μs. **c**, An individual atom phase evolution: δ=20 kHz. **d**, Bloch diagram for **c**. The mark o (x) indicate echo e1 (e2) in **b**. The pulse area of D, R, and RR is $0.5\pi$, $\pi$, and $\pi$, respectively. The marks a-b, c-d, and e-f stand for the atom coherence changes by pulses of D, R, and RR, respectively. The mark 'o' ('x') stands for the timing of echo e1 (e2). Each pulse duration is 0.1 μs. The optical inhomogeneous broadening is 670 kHz (FWHM). $\Gamma_{21}=\gamma_{21}=1$ kHz; $\rho_{11}(t=0)=1$.

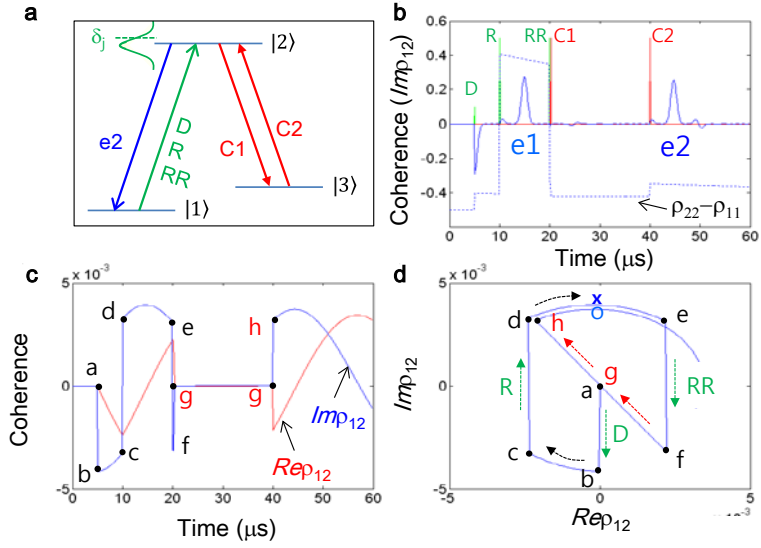

**Figure 6| Controlled Double Rephasing (CDR) echoes. a**, A schematic of CDR, where the control pulses C1 & C2 are added to DD scheme in Fig. 5. Pulses D, R, RR, C1, and C2 arrive at t=5, 10, 20, 20.1, and 40μs. Each pulse duration is 0.1μs. **b**, Numerical simulations of CDR echo in **a**. Each pulse area is π, except D (0.2π). **c, d**, A detuned atom phase evolution for δ=20 kHz. The marks a-b, c-d, e-f, f-g, and g-h stand for the atom coherence changes by pulses of D, R, RR, C1, and C2, respectively. The mark 'o' ('x') indicates echo e1 (e2). The optical inhomogeneous broadening is 670 kHz (FWHM). $\Gamma_{21}=\Gamma_{23}=1$ kHz; $\gamma_{21}=\gamma_{23}=1$ kHz; $\Gamma_{32}=\gamma_{32}=0$; $\rho_{11}=1$. $t_{C1}=32$μs; $t_{C2}=40$μs; $t_{e2}=2t_{RR}-t_{e1}+(t_{C2}-t_{C1})=45$ μs.

**Supplementary Information**

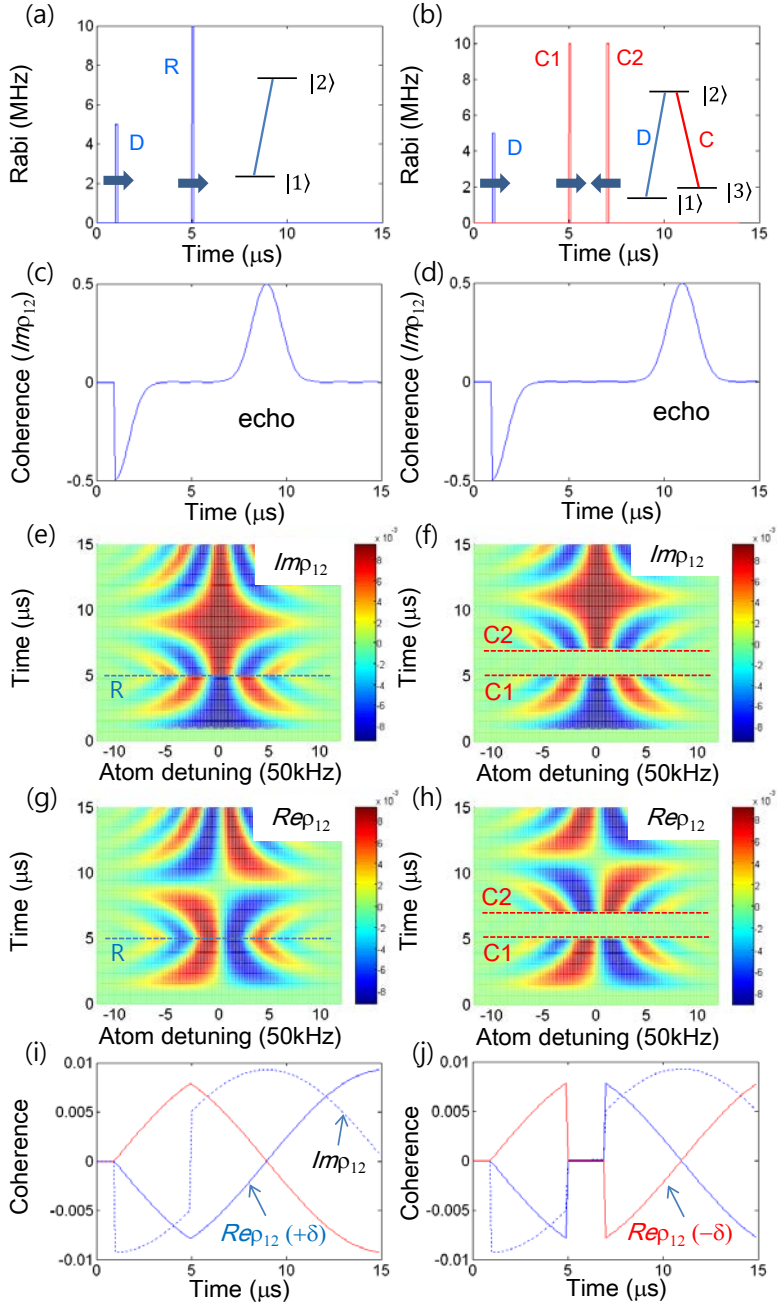

Supplementary Figure 1: Two-pulse photon echo vs. CRIB echo. Schematics of (a) two-pulse, and (b) CRIB echoes. (c), (e), (g), (i) Numerical simulations of two-pulse photon echoes. (d), (f), (h), (j) Numerical simulations of CRIB echoes. The optical Doppler width is 510 kHz (FWHM), where 10 kHz-121 groups are used for the calculations. In (i) and (j), $\delta=40$ kHz. In (j), $Re\rho_{12}(-\delta)$ after C2 (t>7) is the same as in (i) $Re\rho_{12}(+\delta)$ after R (t>5). This represents coherence swapping between symmetrically detuned atom pairs by C1 and C2. All decay rates are zero.

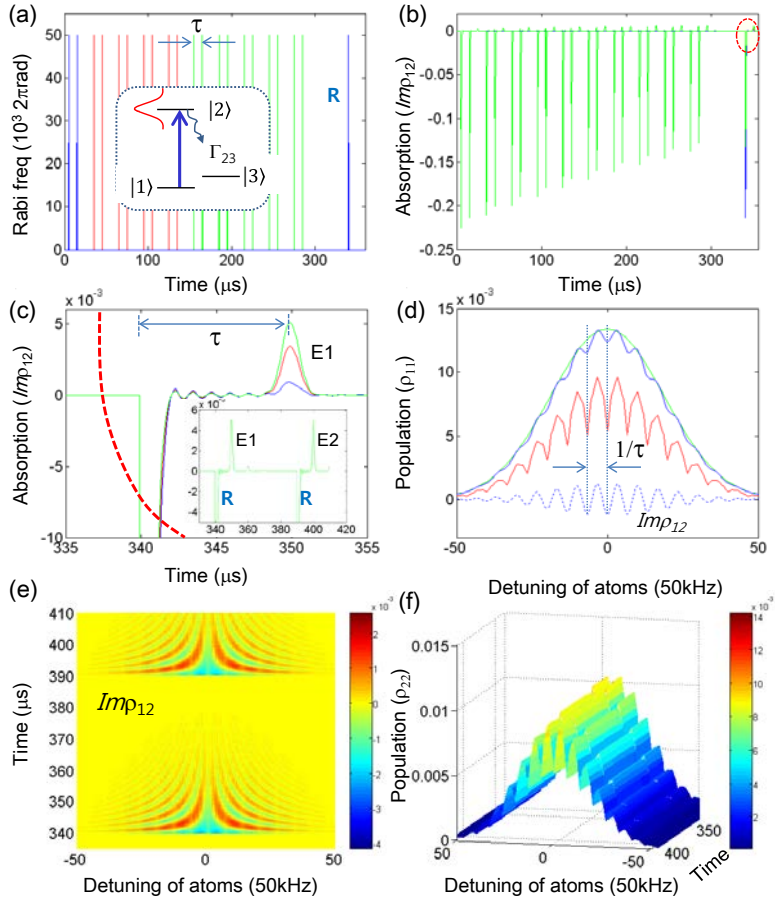

Supplementary Figure 2: AFC echoes. (a) A schematic of AFC echoes. Each pulse duration is 0.1 μs. R is the quantum data pulse (i.e., a read-out pulse). Each two-pulse set whose delay τ is 10 μs arrives at t=5, 35, 65, 95, 125, 155, 185, 215, 245, and 275 μs. The read-out pulse R is on at t=340 μs. The inhomogeneous broadening of the ensemble is 670 kHz (FWHM). $\Gamma_{21}=\Gamma_{23}=10$ kHz; $\gamma_{21}=\gamma_{23}=15$ kHz; All other decay rates are zero. (b) Atomic coherence $Im\rho_{12}$ for (a). Red dotted circle indicates AFC echo. (c) Expansion of the red-dotted circle in (b) for 1- (blue), 5- (red), and 10- (green) sets of two-pulse train in (a). Color matched. Inset: Double read-outs. (d) Population grating $\rho_{11}$ on the ground state |1> created by 1- (blue) and 10- (red) sets of the two-pulse train in (a). Green curve is a reference excited by only the first pulse at t=5.1 μs. Blue-dotted one is a phase grating of $Im\rho_{12}$ at t=15 μs, just before the second pulse. (e) A 2D color map of $Im\rho_{12}$ for the double read-outs in the Inset of (c). (f) A 2D color map of $\rho_{22}$ for (e). The step like reduction in $\rho_{22}$ is used for the echo generations of E1 and E2 in the Inset of (c).

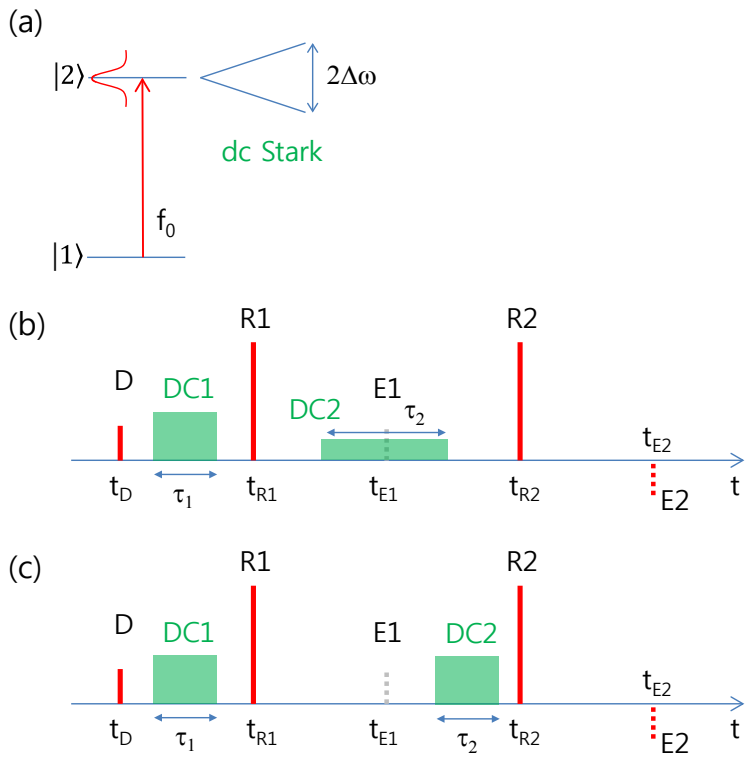

Supplementary Figure 3: Schematics of dc Stark echoes. (a) dc Stark splitting. (b) and (c) Pulse sequence for two different cases of dc Stark echoes (see ref. 74).